# Software Process Improvement Based on Defect Prevention Using Capability and Testing Model Integration in Extreme Programming

Md. Habibur Rahman[1,2] , Ziaur Rahman[1,3] , Md. Al - Mustanjid[4] , Muhammad Shahin Uddin[1] , and Mehedy Hasan Rafsan Jany[1]

[1] Mawlana Bhashani Science and Technology University, Tangail, Bangladesh
mdhabibur.r.bd@ieee.org
[2] Bangabandhu Sheikh Mujibur Rahman Digital University, Kaliakair, Bangladesh
[3] RMIT University, GPO Box 2476, Melbourne, VIC 3001, Australia
rahman.ziaur@rmit.edu.au
[4] Daffodil International University, 102/1 Sukrabad, Dhanmondi,
Dhaka 1207, Bangladesh
mustanjid.se@gmail.com, shahin.mbstu@gmail.com, rafsan.mbstu@gmail.com

**Abstract.** Nowadays, Software Process Improvement popularly known as SPI has been able to receive an immense concern in the continuous process to purify software quality. Several Agile methodologies previously have worked with Extreme programming (XP). Before improving the process, defect prevention (DP) is inevitable. In addition, DP largely depends on defect detection either found earlier in the design and implementation stages or held in the testing phases. However, testing maturity model integration (TMMI) has a crucial aspect in DP as well as process improvement of the software. In particular, when software gets validated by being tested and fixed the defects up, it achieves the maximum capability maturity model integration (CMMI) aiming the process improvement. Here, the article has proposed an improved defect detection and prevention model to enhance the software process following the approach of XP. Besides, as a unique contribution, we have united the capability and testing model integration to ensure better SPI.

## 1 Introduction

The Agile development method is pursued on the basis of customer demands and front planning [1]. Evolutionary learning, incremental development, and Agile methodology inclusion increase customer satisfaction [2]. Small to medium-sized companies rely on Agile's light SPI to endure in the globe of competitive development. SPI in Agile gears up the development team to derive the desires of the organization and prop up software products [3]. This demand encourages researchers to enhance SPI in Agile with stronger accuracy [4]. XP is the most notable software development methodology among several Agile methodologies [5]. XP differs mainly in placing a higher value on adaptability than predictability from traditional methods. XP connects all the practices of the waterfall and iterative model in software development projects [6]. XP prescribes a set of daily procedures from executives and developers. The high expectation within a short time is possible in XP to improve the software process. XP aims to unite humanity and productivity. The customer puts the desirable requirements and the SPI technical team checks the anticipated results based on the perceived criteria of the customer. The ultimate goal of XP is to alleviate modification expenditures [7]. It is undeniable to compromise on quality in the XP. Without DP, software quality cannot be assured. Any inconsistency in the process of software development is classified as a mistake, error or defect [8]. Defects can emerge at any level during the life cycle of software development. For such factors, DP plays a key role in the software system quality refinement. Before making a software, it is best to locate defects. Recognizing defects and triggers of these defects is the most fundamental but mostly overlooked which is essential for quality control [9–11]. DP is a gradual way of finding the root causes of defects and changes the way recurring faults are noticed in SPI [12]. This resistant action cuts development costs and results in higher-quality and reliable software development [13]. Defects must be operated at all steps to preserve customer integrity [14].

Only DP is not adequate to measure SPI empirically. Process management, measurement and other activities need to be concerned along with prevention [13]. CMMI advises organizational interventions, and features to increase product demands, analysis and refurbishment. CMMI software development framework model concentrated on the software development organizations' maturity evaluation [15,16]. CMMI has two parts, one weighing the potential of the process zone as well as the other reflecting organizational maturity at specific levels. CMMI is the SPI initiative that noticeably prevents defects during software development and production life cycle [17]. On the other hand, CMMI is intended to gain high quality by conveying cost optimization, productivity, customer satisfaction and return on investment (ROI) [18]. Further, CMMI requirement implementation guides are conducted by Agile methodology marking the queries which are not marked through the CMMI framework [2].

According to [19], the immature testing process does not efficiently identify defects in SPI. The TMMI is a reference framework and guidance for improving the test process. TMMI attains the concepts of CMMI. In test engineering and development cycle, TMMI uplifts the organization testing process along with software quality and productivity. It is categorized into five stages. The highest satisfaction is mentioned with extremely high maturity level. Satisfying all processes in every maturity level an organization can gain top maturity level step by step [18].

Thus, this paper recommends a model for DP in Agile with XP to deliver SPI more productively. The model standardization is ensured by the combination of CMMI along with TMMI.

The structure of this article is as follows. Section 2 is provided with related works. Section 3 illustrates the proposed model. Section 4 analyzes the results meet up with model. Section 5 concludes the paper with future work.

## 2   Related Works

The effectiveness of CMMI in the development of Agile software was evaluated in [20] based on a systematic review of the literature published up to 2011. Several criteria were used to measure CMMI adeptness in Agile development. The study also determined that CMMI can be achieved using Agile software development.

The article [17] explored how SPI treats defects of software development. Comparisons between the scrum and XP model explored in [7] placed by their model features that lead to a deep understanding of these models. Authors in [21] deployed an improved model with consideration of XP's weak documentation and architectural design. A new framework tailored by integrating testing phases at each step to tackle the deficiency of the XP criteria outlined in [22].

Authors in [3] investigated 423 detailed articles in consonance with SPI aspects published from 2001 to March 2013. This paper also provided a systematic review of literature on the SPI fields mainly studied in previous research. The research found that new SPI techniques need to be built which really compares traditional techniques to run Agile more fluently. Authors concluded that far more research on Agile strategies is required.

Studies above clearly point out that DP in SPI of Agile methodology, particularly in XP requires a new initiative for progression. Moreover, this paper suggests a model to refurbish SPI through DP in XP. SPI and quality of the software are quantified by the incorporation of CMMI with TMMI.

## 3   Proposed Model

DP tasks are consistently investigated; these are rolled out by each of the allocated teams. Predetermine the causes of defects as it is required to fully overcome the consequences of defects. Addressed action items and set priorities focused on a causal interpretation throughout the assessments. Thereafter, to prevent the defects, the expense of implementing structural changes is estimated. Finally, the intended influence on software quality is taken into account. Figure 1 signifies the DP strategy of software development process. The critical role in the execution phase is to develop a plan of action. At first, members of the software development group meet to plan for the duties and relevant DP operations. There is a kick-off meeting held to convey information about the implementation process among members of the team. The session outlines the procedures, standards, protocols, techniques, and tools appropriate to the functions of the software. Developers concentrate on the latest modifications based on vibrant specifications. Assumed outcomes and evaluating approaches for evaluation are analyzed. Generic errors and advised corrective efforts are included; and also team assignments, a task scheduler and project targets are instituted.

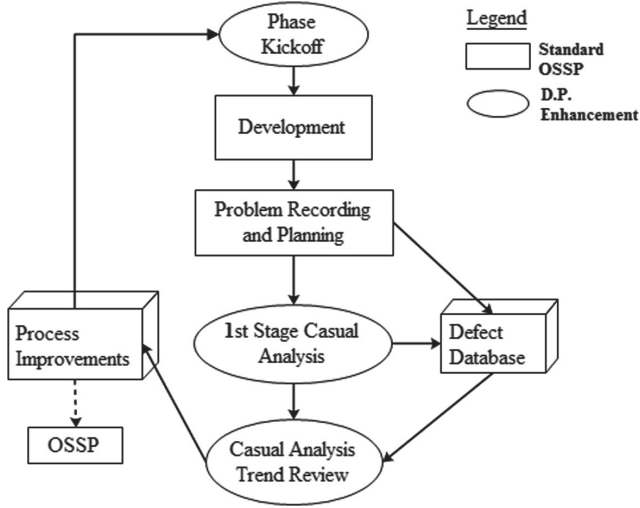

**Fig. 1.** Defect prevention strategy for software development process; Here in the Figure, OSSO stands for organization standard software process

Management in the IT sector must be committed to practicing a written DP policy at both the organization and project stage presented in Fig. 2. In order to strengthen software processes and items, long-term plans should be adopted for funding, resources and implementation. In Fig. 2 the defect detection and prevention are described in XP. Developers must pay attention to client stories determining their values with all along software development life cycle (SDLC). The acceptance test standards should also be fixed in the iteration plan. In the planning phase, these are conducted. The design has been done based on the software development plan. Coding with pair programming and refactoring are begun after the design phase. Refactoring is accomplished on the basis of the defects encountered in the test stage. Utilizing unit testing, programmers track the output. Testing continues with ongoing integration shortly after the coding. In this stage, the entire acceptance along with the user acceptance test is executed.

According to the SDLC, a whirlpool diagram in Fig. 3 is proposed and tested for DP on a small project. Customers always prefer to satisfy their requirements into 100%. In practice, it is quite impossible to develop that kind of software. Thus, we propose a model in Fig. 3 provides more than 90% defect free software and achieved the highest maturity. In this model, perfection is boosted in anti-proportional rate by limiting the software defects. Hence, after several phases, the model optimizes the defects in the downward cycle.

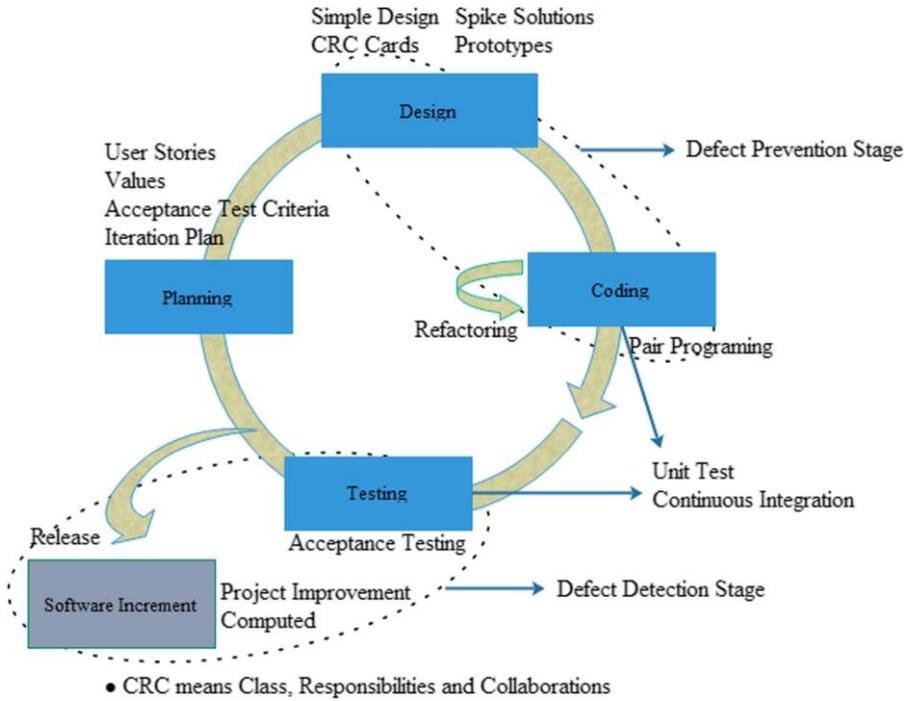

**Fig. 2.** Defect detection and prevention model in Extreme programming

## 4 Result Analysis

There are many software process evaluation methods and norms [23,24]. DP is gradually improving the process of production. The preceding Eq. (1) ensures software quality.

$$QA = QC(Defect\ Identification) + DP \quad (1)$$

where, QA = Quality Assurance; DP = Defect Prevention and QC = Quality Control

Standard quality is upheld using parameters for judging the quality of the software. To enhance the software mechanism and acquire qualified software, the following eight substantial measurement parameters are regarded.

Here,
Q1 = Integrity
Q2 = Correctness
Q3 = Maintainability
Q4 = Efficiency
Q5 = Reliability

Q6 = Modifiability  
Q7 = Reusability and  
Q8 = Portability  

Integrity measures code and information consistency by monitoring an authorized individual's access to software. Correctness examine the expectations of the user in software and how certain requirements are. Maintainability suggests how much effort will be made to address and follow up defects that occurred in the operational processes of software. Efficiency gets to decide what further computing logistics and code a program requires to deliver a function. Reliability ensures that the expected function of a program is performed with the appropriate clarification while it ought to be performed. Especially in packaging and function scopes, a software program performs as needed in another application helps to ensure reusability. The term modifiability defines the adjustment capability of the additional features of the customer requirements. Another term portability indicates the platform independence of the software [25].

A testing and capability maturity model as leveled in Table 1. If the software satisfies the integrity and correctness, the capability maturity will be initiated. The software capabilities will be characterized by modifiability, reusability, and portability. Containing all the quality parameters, the testing maturity will be high, and the ability will be scalable.

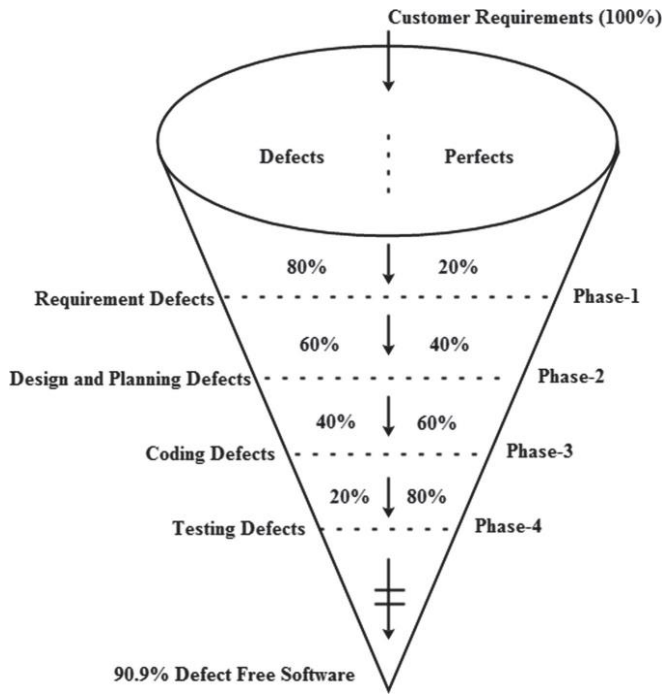

**Fig. 3.** Software defect prevention whirlpool diagram in Extreme programming in Agile methodology

**Table 1.** Software process improvement with quality parameters in the CMM and TMM integration

| CMM level | Focus | Satisfaction of TMM | Quality parameters |
| --- | --- | --- | --- |
| 5. Optimizing | Continual SPI | Extremely high | Q1, Q2, Q3, Q4, Q5, Q6, Q7 and Q8 |
| 4. Managed | Performance and software quality both are measured | High | Q1, Q2, Q4 and Q5 |
| 3. Defined | Well organizational process and support | Medium | Q6, Q7 and Q8 |
| 2. Repeatable | Software management process | Low | Q3 and Q6 |
| 1. Initial | Ad-hoc and unorganized | Extremely low | Q1 and Q2 |

**Table 2.** Maturity level of defect detection and prevention

| TMM (Detection) | CMM (Prevention) | |
| --- | --- | --- |
|  | *Optimized* | *Managed* |
| Extremely high | Highly matured (++) | Matured (+×) |
| High | Matured (×+) | Immature (××) |

Table 2 and 3 represented the maturity and acceptance level respectively. In Table 2, the maturity is considered high when the capability fulfills the optimized level. If the defects are high and avoided at the manageable level, it will be acknowledged immature otherwise it will be matured. Levels of acceptance just set out in Table 3. If the amount of recognition of defects is poor and occurred continuously, it will be discarded. Whereas the identification is medium and can be specified by the model, it is considered acceptable or negotiable.

**Table 3.** Acceptance level of defect detection and prevention

| TMM (Detection) | CMM (Prevention) | |
| --- | --- | --- |
|  | *Defined* | *Repeatable* |
| Medium | Acceptable($\oplus \oplus$) | Negotiable ($\oplus \otimes$) |
| Low | Negotiable ($\otimes \oplus$) | Rejected ($\otimes \otimes$) |

By implementing the proposed model on a small project, we achieved 90.9% defect free software [https://github.com/Al-mustanjid/Travellers_Project]. Figure 4 portrays the outcomes of our overall approach. We spotted more than 80% defects in the requirement specification. Forward to passing planning, coding and testing phases, we lower the defects to 20% throughout the testing phase. Software is launched for a short period of time after that alpha version.

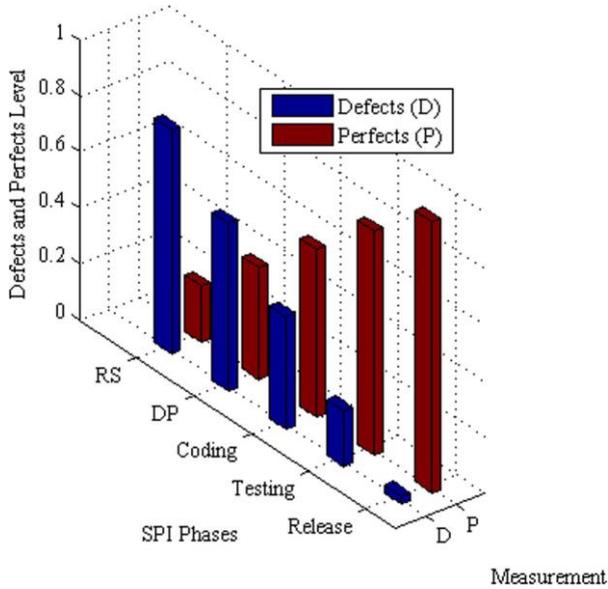

**Fig. 4.** Expected defects reduction by TMMI Phase; RS stands for Requirement Specification and DP is for Design and Planning

The faults are decreased to more than 90% since obtaining the reviews and customer feedback in circular phases. Finally, we got the highest degree defect free software implemented in XP.

## 5 Conclusion

XP is one of the most widely used methodologies for small software project development where DP is inevitable towards ensuring quality software delivery. As per our investigation, no defect detection and prevention model in SPI that in particular can solve issues encountering by the industry could desirably reduce the defects step by step in the Agile. As capability based CMMI and testing based TMMI approach measure software standardization, but proposing a novel approach integrating both could envisage a new dimension to encourage industry practicing Agile XP.

In this paper we have observed through the experiments that proposed model applied to the software projects can significantly reduce defects as clarified by the relevant sections. To fulfill the software quality improvement standards, this proposed model is able to achieve the highest level of optimization and detection that can also be considered as a separate contribution to the authors. The future includes experimenting the model with large scale project implementation.